\documentclass[manuscript]{aastex}

\def\gtorder{\mathrel{\raise.3ex\hbox{$>$}\mkern-14mu
             \lower0.6ex\hbox{$\sim$}}}

\def\ltsima{$\; \buildrel < \over \sim \;$}
\def\simlt{\lower.5ex\hbox{\ltsima}}
\def\gtsima{$\; \buildrel > \over \sim \;$}
\def\simgt{\lower.5ex\hbox{\gtsima}}

\begin{document}


\title{Photometric Identification of Young Stripped-Core 
Supernovae}


\author{Avishay Gal-Yam\altaffilmark{1}}
\affil{Astronomy Department, California Institute of Technology,
Pasadena, CA 91125}
\email{avishay@astro.caltech.edu}
\author{Dovi Poznanski, Dan Maoz}
\affil{School of Physics \& Astronomy, Tel-Aviv University, Tel-Aviv
69978, Israel}
\email{dovip@wise.tau.ac.il, dani@wise.tau.ac.il}
\and
\author{Alexei V. Filippenko, Ryan J. Foley}
\affil{Deaprtment of Astronomy, 601 Campbell Hall, University of California, Berkeley, CA 94720-3411}
\email{alex@astro.berkeley.edu, rfoley@astro.berkeley.edu}


\altaffiltext{1}{Hubble Fellow.}


\begin{abstract}

We present a method designed to identify the spectral type of young
(less than $\sim30$ days after explosion) and nearby 
($z \simlt 0.05$) supernovae (SNe) using their broad-band 
colors. In particular, we show that stripped-core SNe (i.e.,
hydrogen deficient core-collapse events, spectroscopically
defined as SNe~Ib and SNe~Ic, including broad-lined
SN 1998bw-like events) can be clearly distinguished
from other types of SNe. Using the full census of nearby
SNe discovered during the year 2002, we estimate the impact 
that prompt multi-band photometry, obtained by $1$~m class 
telescopes, would have on the early identification of 
stripped-core events. Combining this new approach with
ongoing spectroscopic follow-up programs, one can expect
$\sim 20$ nearby, stripped-core events to be identified, each year,
around, or before, maximum light. Follow-up studies, including 
prompt, multi-epoch optical spectroscopy and spectropolarimetry, as well 
as radio and X-ray observations, could greatly increase our understanding 
of these events, and shed new light on their
association with cosmological gamma-ray bursts.

\end{abstract}




\section{Introduction}

Several types of supernovae (SNe) are recognized, and are
most commonly determined by the attributes of the SN
optical spectra. The most basic division is between 
hydrogen-rich (type II) SNe, and hydrogen deficient
(type I) SNe (Minkowski 1941). Type I events are
further divided into SNe Ia, which usually show 
the prominent Si II absorption trough near $6150$~\AA,
SNe Ib, which lack this feature but show prominent 
He lines, and SNe Ic, whose spectra show neither
(see Filippenko 1997 for a review).

Direct detection of the progenitor stars on pre-explosion
images has established that type II SNe result from the
explosion of massive stars, presumably following the
gravitational collapse of the stellar core (SN 1987A, White \& Malin 1987;
SN 1993J, Aldering et al. 1994; SN 2003gd, Van Dyk, Li, \&
Filippenko 2003, Smartt et al. 2004). The existence
of transition objects, which appear to be SNe II in spectra
obtained shortly after explosion, but later evolve and become
very similar to SNe Ib and SNe Ic (SNe IIb; e.g., Filippenko 1997), 
as well as the apparent association of SNe Ib and Ic with young stellar
populations (e.g., Van Dyk, Hamuy, \& Filippenko 1996), suggest that these events also result
from the core collapse of massive stars. These stars are assumed to have
lost most (in the case of SNe IIb) or all (in SNe Ib and Ic) of their hydrogen 
envelope, as well as, in the case of SN Ic progenitors, most of their
helium, either through stellar winds or due to binary interaction.  
In contrast, SNe Ia are frequently found among old stellar populations, e.g., in
elliptical galaxies, and are unlikely to be associated with very young
progenitor stars. These events are commonly assumed to be the result of
a thermonuclear runaway explosion of a white-dwarf star, which reaches
the Chandrasekhar mass either through accretion from, or a merger with,
a binary companion.

Of all the various types of SNe, the ones that are probably the least
studied are SNe Ib and Ic, which will be referred to henceforth as
stripped-core events. Contributing factors probably include the intrinsic 
rareness, relatively low luminosity, and heterogeneity of these events, 
along with the absence of a high-profile science driver for such work, 
as provided, for example, by cosmological distance estimation for studies
of SNe Ia. Most of the progress in this field was due to the (rare) 
occurrence of bright stripped-core events in nearby galaxies, e.g., SN 1994I
in M51. The only recent comprehensive study of stripped-core events was
conducted by Matheson et al. (2001).
 
This situation is perhaps about to change, following the emerging connection 
between stripped-core SNe and long-duration gamma-ray bursts, and in particular, the associations between 
GRB 980425 and SN 1998bw (Galama et al. 1998) and GRB 030329 and SN 2003dh 
(Stanek et al. 2003; Hjorth et al. 2003; Matheson et al. 2003) -- see Lipkin et al. (2003)
for a recent review. The amount of overlap between GRBs and stripped-core SNe 
is intriguing. Are all GRBs associated with a SN? Are
all stripped-core SN explosions triggered by, or associated with,
a GRB or GRB-like mechanism (i.e., a non-isotropic, jetted explosion;
Khokhlov et al. 1999) ?
Early-time studies of stripped-core SNe may provide the answer to this
last quesion.

\section{The Importance of Early-Time Identification
of Stripped-Core SNe}

A key to understanding the connection between stripped-core
SNe and GRBs is whether ``ordinary'' SNe Ib and Ic (such as SN 1994I)
and SN 1998bw-like events, associated with GRBs, are bridged by a 
continuum of SNe with intermediate properties. Alternatively,
are there two distinct populations of SNe -- those associated
with GRB-like explosions, and ``ordinary'' stripped-core events. The existence
of a continuum would support theories in which GRB-like phenomena
(i.e., highly aspherical explosions) occur in all stripped-core
events, with the observed variety driven by geometrical (orientation)
effects, or variations in the intrinsic energy of the central GRB 
engine, in which case ordinary SNe are ``failed GRBs.'' A bi-modal
distribution would suggest that the (mostly spherical) core-collapse
and the (highly aspherical) GRB are two unrelated physical processes that
occur simultaneously in some cases (e.g., Soderberg, Frail, \& Weiringa 2004).  

Since the spectral distinction between SN 1998bw-like events and
normal SNe Ic becomes less pronounced with age, it is quite possible
that the fact that most stripped-core events are spectroscopically
observed at or after maximum brightness causes us not to detect
intermediate events, which resemble SN 1998bw early on, but are already
ordinary-looking when we observe them. Thus, early-time identification of
stripped-core events is crucial in order to trigger
spectroscopic studies of these events, as well as spectropolarimetric
observations, the most sensitive tracers for explosion asphericity.

The collapsar model predicts an association between long-soft GRBs and stripped-core
SNe (Woosley 1993, MacFadyen \& Woosley 1999).  The progenitor stars are required 
to have lost their extended hydrogen-rich envelopes before collapse, 
either to a binary companion or through mass loss by stellar winds. This is necessary for 
a relativistic jet to escape the surface of the star on a collapse timescale for the 
helium core ($\sim10$~s). 
The SNe Ic associated observationally with GRBs (SN 1998bw
and  SN 2003dh) are luminous, with models indicating large explosion energies ($\sim10^{52}$ erg) 
and the production of large masses of radioactive Ni ($\sim0.2-0.5 M_{\odot}$) (Iwamoto
et al. 1998; Woosley, Eastman, \& Schmidt 1999; Mazzali et al. 2003; Woosley \& Heger 2003).  
Collapsars can produce large amounts of Ni by expelling hot material 
from a poorly cooled accretion disk (MacFadyen 2003) 
around the central accreting black hole powering the GRB.
Current models incorporate asymmetry in the explosion to explain the fast rise in the light 
curve of SN 2003dh (Woosley \& Heger 2003). Thus, 
early photometric observations of stripped-core SNe on the rise to maximum 
will better constrain the degree of asymmetry in stripped-core SNe, and,
in combination with early spectra, also the 
explosion energy and nickel masses. These quantities are critical to understanding 
the mechanism responsible for exploding the star and, sometimes, producing a GRB.  

SN 1998bw was uniquely radio luminous. In spite of recent and ongoing
intensive efforts (Berger et al. 2003; Soderberg et al. 2004) no
similar events were discovered among $\sim 60$ stripped-core SNe observed
so far (1999--2004).  On the other hand, the late
initiation of many of the radio searches may have contributed to this null
result. In particular, the radio light curve of SN 1998bw
peaked early, $\sim~5$ days {\it prior} to optical maximum. 
Another SN 1998bw-like event, SN 2002ap,
one of the only other detections, also peaked early
in the radio band. However, with a radio luminosity $10^4$
times below that of SN~1998bw, this event was only detected
since it is the nearest SN in the sample ($\sim 7$ Mpc) and
was observed $\ge 10$ days before optical maximum.
Thus, early identification of stripped-core events would be highly
valuable for radio studies of these SNe. The same is true in the case of
space-based searches for the rapidly decaying X-ray emission from these
events.

To conclude, early identification of stripped-core events, providing
a trigger for detailed follow-up observations, would be very helpful 
in characterizing the little-known properties of the stripped-core 
SN population, as well as in searching for transition objects between GRB-associated SN 1998bw-like
events and more ordinary stripped-core SNe. Such studies would further
our understanding of stripped-core SN physics, and provide an important
insight into the GRB-SN connection. 
 
\section{Color Classification of Young Supernovae}

In Poznanski et al. (2002) we describe a method to classify
SNe based on their broad-band colors. The method is based on
a large number of high-quality optical SN spectra we have assembled (Table 1). 
This database includes observations of the prototypical members of 
every well-defined SN subclass, at various ages.
By redshifting the spectra and convolving them with the appropriate
filter throughput curves, we derive synthetic photometry of SNe
of all types at a given redshift, which can then be compared to
available measurements, in order to determine, or constrain,
the possible type of an observed SN. The tools we developed
are quite general, and can be used to classify SNe of
all the well-defined spectroscopic subclasses at arbitrary
redshifts. Observations in optical bands may be used to classify
SNe up to a redshift of $z=0.75$, while infrared observation
are required at higher redshifts. 

In view of the growing interest in broad-lined
SN 1998bw-like events, following their established 
association with GRBs, we have added the spectra of the two 
best-observed examples, SN 1998bw (Patat et al. 2001)
and SN 2002ap (Gal-Yam, Ofek, \& Shemmer 2002; Foley
et al. 2003) to our spectroscopic database (Table 1)\footnotemark. 
\footnotetext{An updated version of our web-based
SN-typing tool, incorporating these additional spectra,
is now available at \url{http://wise-obs.tau.ac.il/$\sim$dovip/typing}}  
We then 
proceed to study in detail the case of nearby ($z \simlt 0.05$)
and young (less then $\sim$30 days after explosion) 
SNe. As we show below, young stripped-core SNe
(Ib and Ic, including the SN1998bw-like events) can
be clearly distinguished from all other types of SNe.

The broad-band colors of SNe are determined by their
spectral energy distribution, dominated by the continuum
shape. There are many spectroscopically defined 
subtypes of SNe. However, some of
the more subtle spectroscopic divisions are erased when
we examine only the continuum shape. Considering only 
SNe near peak brightness, we find that all subtypes
of hydrogen-rich SNe (types II-P, IIn, and IIb) have similar
spectra, rising steadily toward the blue all the way 
down to the atmospheric cutoff at $\sim3000$~\AA. 
Type Ia SNe are generally bluer than SNe~II at wavelengths
longer than $4000$~\AA, beyond which a UV cutoff, attributed to
line-blanketing by iron-group elements, strongly
suppresses the flux (see Riess et al. 2004 for a relevant 
discussion). Type Ib and Ic SNe near maximum have spectra 
similar to each other,
typically redder than both SNe~Ia and SNe~II, and showing a 
flux deficit blueward of $\sim5500$~\AA. 

One of the peculiarities of SN 1998bw was the relatively blue
spectrum observed at early times, significantly bluer than 
most other SNe~Ic. This is demonstrated 
in Figure 1, showing the rest-frame spectra of SN 1998bw compared
with three other nearby SNe observed close to peak brightness. 
Other SN 1998bw-like events, such as SN 2002ap,
also have somewhat bluer continua, intermediate between those
of SN 1998bw and those of ``normal'' SNe Ic and SNe Ib.
 
Inspecting Figure 1 we can see that young SNe of all types
have similar continuum shapes redward of $\sim5500$~\AA. 
We are thus led to focus on bluer bands in our search for 
broad-band classification schemes for young SNe. Experimenting
with various choices of bands, we find that $B-g$ vs. $V-R$\footnotemark
\footnotetext{Throughout, we note the Johnson-Cousins 
and SDSS filter systems by $UBVRI$ and $ugriz$, respectively; see Poznanski 
et al. 2002 and references therein for exact details.} 
plots proves the most informative. The enhanced sensitivity
one would expect to gain by analyzing also
$U$-band data is in fact hindered by poor existing spectroscopic 
coverage of young SNe, resulting in sparse sampling of
the relevant color spaces by our spectral database. The
observational difficulties expected in performing accurate
$U$-band photometry are likely to further impede the
utility of such data. We therefore consider below mainly
$B-g$ vs. $V-R$ plots.
 
\begin{figure*}
\includegraphics[width=17cm]{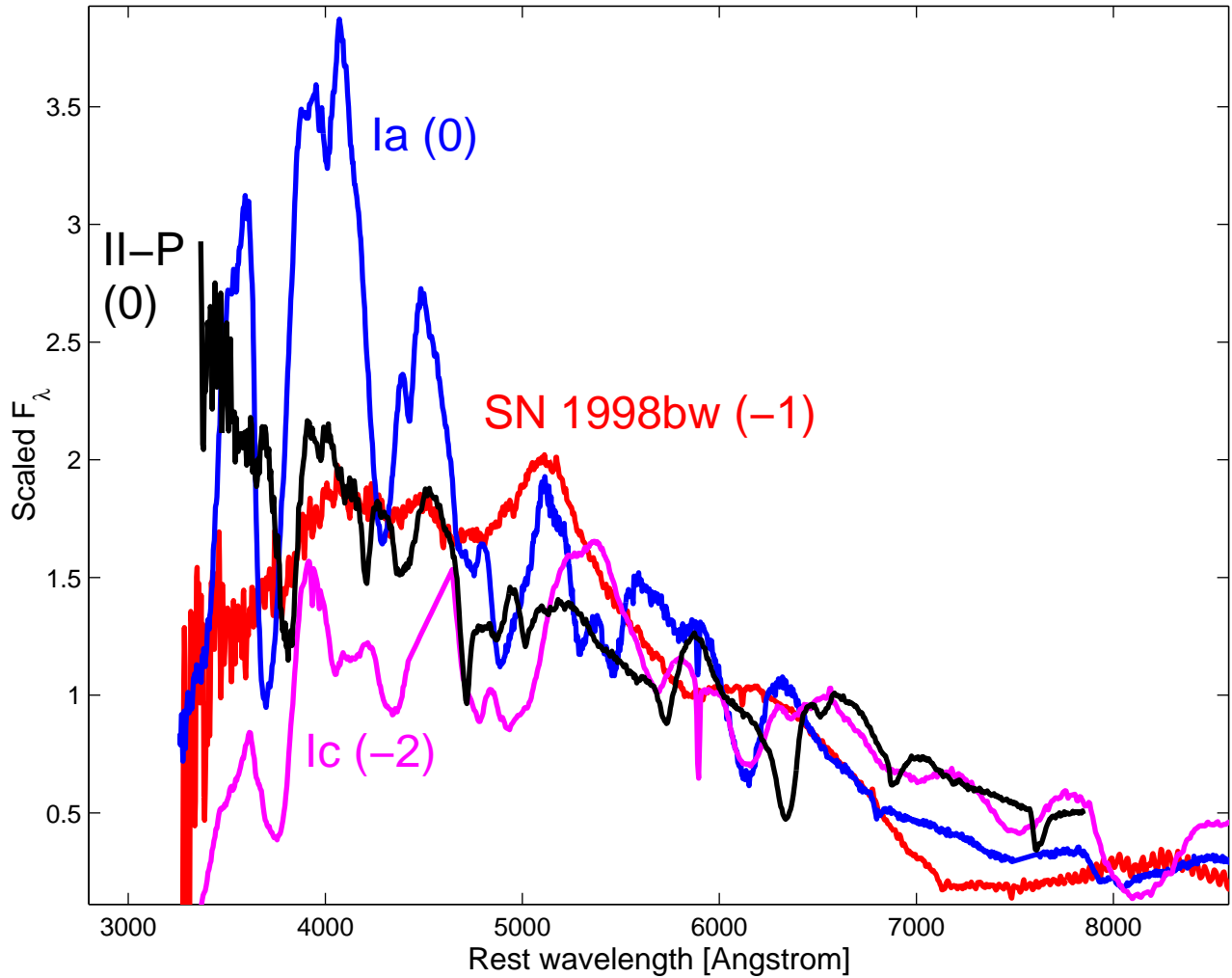}
\caption{Spectra of SNe near maximum light. For clarity,
we show only spectra of the prototypical SN~Ia 1999ee 
(blue; Hamuy et al. 2002), SN~Ic 1994I (Magenta; from Filippenko et al. 1995), SN II-P 
1999em (black; Hamuy et al. 2001),
and SN 1998bw (red; Patat et al. 2001), with ages in days
with respect to $B$-band maximum marked. All spectra have been  
normalized to have the same flux around $6000$~\AA. Spectra of other
types of SNe~II (e.g., SN IIb 1993J or SN IIn 1998S) are similar
to that of SN II-P 1999em, while those of SNe~Ib are similar 
to those of the SN Ic 1994I. SN 2002ap is intermediate between 
SN Ic 1994I and SN 1998bw. 
Note how all SNe have similar spectral shape redward of 
$5500$~\AA, thus making bands redder than $V$ less useful for
classification of young SNe.} 
\end{figure*} 

Figure 2(a) shows the $B-g$ vs. $V-R$ color-color diagram
for SNe of all types up to $\sim1$ week after peak brightness.
This and following figures were produced using the 
methods discussed in detail by Poznanski et al. (2002).
A color cut at $B-g=0.24$ mag separates stripped-core SNe
from other SNe (SNe~Ia and SNe~II). Considering only events with
$B-g>0.24$ mag, stripped-core SNe form a sequence
in $V-R$ colors, going from the bluest SN 1998bw, through
SN 2002ap and SNe~Ic, with SNe~Ib being the reddest. 
Similar analysis can be carried out for SNe at
higher redshifts. Figure 2(b) shows that for SNe 
at $z=0.05$, the $g-V$ vs. $V-R$ color-color plot can be 
used for classification with similar efficacy.

Note that the use of these plots requires colors measured with a 
photometric accuracy of $\sim0.05$ mag, or better. As we further discuss below,
for nearby, $z\le0.05$ stripped-core SNe, this level of accuracy can
be easily obtained with $1$~m class telescopes. The plots also show the
possible effects of dust reddening. For example, examining Fig. 2(a), we
see that young SNe of any type are unlikely to be confused with SN 1998bw
or SN 2002ap-like events, even when strongly affected by dust. Some reddened
SNe II may be confused with young SNe Ib or Ic, though. Another encouraging
fact is that SNe Ia, which, being the most luminous type of SNe, usually dominate
flux-limited SN surveys, are unlikely to appear similar to stripped-core events,
even if highly extinguished ($A_V > 1$ mag).  

\begin{figure*}
\includegraphics[width=7cm]{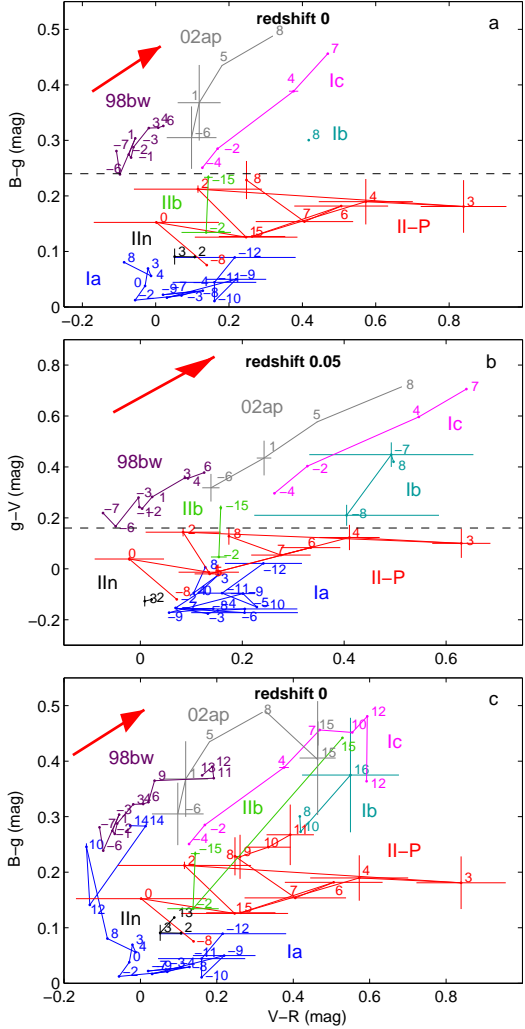}
\caption{{\bf (a)}~$B-g$ vs. $V-R$ colors for SNe of
all types, up to $\sim1$ week past maximum
light. SN ages (in days) with respect to $B$-band maximum
light are marked. A color-cut at $B-g=0.24$ mag separates
stripped-core SNe from all other SNe. The error bars
represent extrapolation uncertainties introduced when
spectra do not cover the entire wavelength range
of a given filter. The arrow shows the reddening effect
corresponding to $A_V=1$ mag of dust extinction. 
See Poznanski et al. (2002) for a full discussion.
{\bf (b)}~$g-V$ vs. $V-R$ colors for SNe up to $\sim1$ week 
past maximum light, but for SNe at $z=0.05$. 
A color-cut at $g-V\sim0.16$ mag separates stripped-core SNe 
from most other events. {\bf (c)}~$B-g$ vs. $V-R$ colors for SNe of
all types, up to $\approx$ two weeks past maximum
light. Note that SNe of type II-P, IIb, Ic and Ib
attain approximately the same colors $\approx2$ weeks
after maximum.} 
\end{figure*} 
 

Classification of SNe becomes harder as they become
older. Figure 2(c) shows $B-g$ vs. $V-R$ 
color curves computed for SNe of all types, at $z=0$,
up to $\sim2$ weeks past maximum light. We can see 
that SN 1998bw can still be easily distinguished from all
other types of SNe. While SNe~Ia, between $10 - 14$ days
past maximum, have similar colors to those of SN 1998bw
before maximum light, elementary photometric information 
(i.e., if the SN is rising or declining in brightness)
will differentiate between these two options. We note that
highly reddened SNe Ia (with $A_V > 1$ mag), 
around two weeks after peak brightness,
could perfectly mimic SN 1998bw, both in color and in 
photometric behavior. However, SNe Ia are usually detected
in relatively dust-poor environments, so such events are 
very rare. Type II-P and IIb
SNe during their second week past maximum have similar
colors to SNe~Ib and Ic and also to those of SN 2002ap
at similar ages ($B-g \approx V-R \approx 0.4$ mag). 
Thus, these types cannot be disentangled using these colors alone.  


\section{Young, Nearby SNe -- Current Status and
Potential Improvement}

In order to utilize the method described above for prompt
identification of stripped-core SNe, one needs to secure
accurate, multi-band photometry of each new nearby SN,
as soon as possible after discovery. Stripped-core SNe of
all kinds, close to peak brightness, are expected 
to be brighter than $\sim16$ mag ($\sim20$ mag) at $z=0.01$ ($z=0.05$)
in the $B$, $g$, $V$ and $R$ bands.
Using the web exposure-time calculator for the $1.54$m Danish
telescope at ESO as an example, we find that a four-band 
photometry sequence for a $20$ mag source could be obtained
in less than $10$ minutes ($35$ minutes) with $S/N$ of $20$ ($50$).
Thus, photometry at the required level of accuracy, as
specified above, could be easily obtained with a $1$~m class
telescope. Since less than a single suitable young SN per day is 
currently discovered, an early photometric classification
program could be maintained at such a facility using only a small
fraction of the available telescope time.

To assess the possible impact of such a prompt 
color-classification program, we inspect a SN sample
comprised of all nearby ($z \simlt 0.05$) SNe detected 
and reported in the IAU Circulars during the year 2002. We
further limit our sample to SNe for which
the reported photometry can be used to constrain the
date of explosion. Specifically, we have compiled all those events
for which the age at discovery could be constrained by
pre-explosion images to be less than $\sim1$ month.
Considering SN rise times (i.e., the time between
explosion and peak brightness, typically between 
one and three weeks) and the fact that most of these
SNe exploded some time after the reported non-detections,
we estimate that this sample is comprised mostly of SNe
that are within one week from peak brightness. We have
to select young SNe using an upper limit on the time since
explosion, instead of their age relative to peak brightness
(as done in the previous section), since the peak date is usually not reported
in the IAU Circulars, and is often unknown or poorly constrained.  

In Figure 3 we plot, for the SN sample described above (blue) and
for stripped-core events only (red), the distribution 
of delay times between SN discovery and the date of the first
spectroscopic observation. As this histogram shows, spectra of only about
half of the SNe are obtained promptly (within $\sim5$ days).
Spectroscopic observations of the rest of these ``young'' SNe (i.e., which were
discovered at an early age) are unfortunately secured only at much later dates.
Comparing the blue and red histograms, we see that the distribution
of delays between SN discovery and 
spectroscopic follow-up for stripped-core events is the same as
the distribution found for the general SN population, as can be expected
in the case where no priors are used in the selection of events for
spectroscopic follow-up. Similar results were found in an analysis 
of SNe reported in IAU circulars in the last 5 years (1998-2002) by
E. Cappellaro (2004, private communication).
It should be noted that the delay times
shown in Fig. 3 are between the date of discovery and the date the
first spectrum was obtained. In reality, further delay (of typically a few days)
in the study of interesting events is caused by the time required to analyze
the spectra, report them to the IAU, and for the IAU to distribute
this information. 

This analysis demonstrates the potential contribution of early
photometric identification to the studies of SNe in general, and
stripped-core SNe in particular. Of the $19$ stripped-core events
discovered probably within 7 days from peak magnitude during the year
2002, only $11$ were identified as such within 5 days from discovery,
using the combined spectroscopic resources of the entire SN-research 
community, which are unlikely to be significantly expanded in the coming
few years. Thus, studies such as the ones described in $\S~2$ were
limited to only $58\%$ of those stripped-core SNe which happened to be
discovered early. Implementing the photometric identification 
procedures outlined in $\S~3$ using a $1$~m class telescope could have
practically doubled the sample of objects available for this type of research.

\begin{figure*}
\includegraphics[width=17cm]{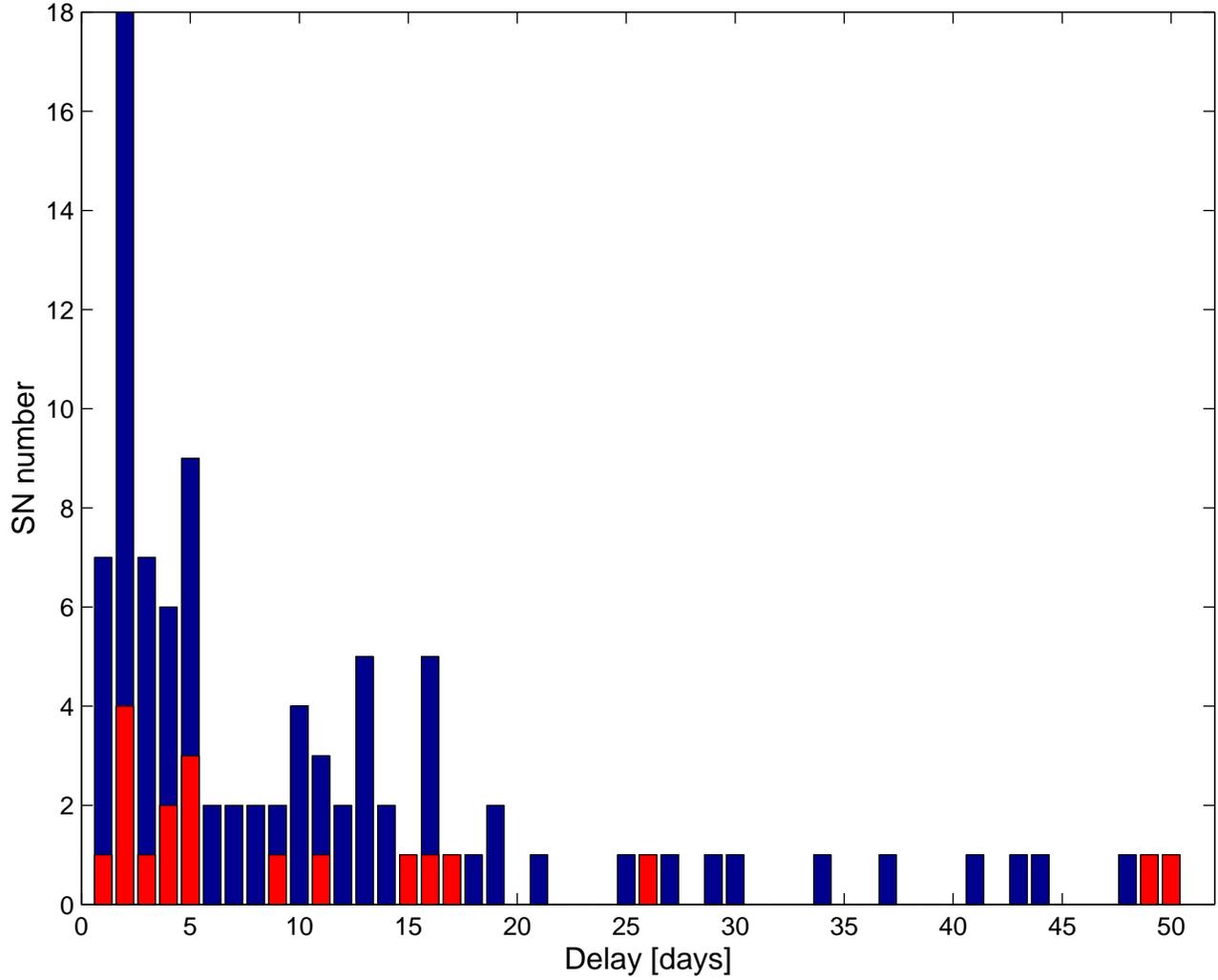}
\caption{The distribution of delay times (in days) between the
SN discovery and the date of the first spectroscopic observation,
for all ``young'' SNe (see text) discovered during 2002 (blue)
and for the subset of stripped-core SNe (red).} 
\end{figure*} 

\section{Conclusions}

The emerging connection between stripped-core SNe and GRBs
is likely to revitalize the study of this class of SNe. The
rareness and relative faintness of these events has so far
held back advancements in systematic studies of their
properties, which would probably help
understanding their relation to GRBs and the underlying
physics. In particular, studies possible only
while these SNe are young (i.e., at, or preferably before,
maximum light) have been hindered by frequent delays
between the discovery, spectroscopic type determination, and
the announcement of the type in IAU circulars. 

To facilitate prompt classification of young stripped-core SN candidates using
broad-band colors, we have added the spectra
of SN 1998bw, associated with GRB 980425, and the similar SN 1998bw-like
SN 2002ap, to the spectroscopic data base of Poznanski et al. (2002). 
We have applied the methods of these authors to this expanded data base,
and have studied in detail methods to classify young (around peak, or before)
and nearby ($z\le0.05$) SNe. We have shown that stripped-core events
can be efficiently selected using four-band photometry focused
on the bluer bands. We have also demonstrated that the required photometric accuracy
can be obtained with a $1$~m class telescope, and that reddening effects
do not significantly compromise our analysis. Application
of this method to the SN population discovered by current SN surveys will probably
double the number of young stripped-core SNe identified at or before
peak magnitude, and, combined with current SN spectroscopy programs,
will supply about $20$ new targets each year for
detailed follow-up studies in radio, optical, and X-ray bands. 
  
\section*{Acknowledgments}

D. Fox, A. MacFadyen, F. Patat and A. Soderberg
are thanked for help, advice, and data used in this work.  
A.G. acknowledges support by NASA through Hubble Fellowship grant
\#HST-HF-01158.01-A awarded by STScI, which is operated by AURA, Inc.,
for NASA, under contract NAS 5-26555. D.M. acknowledges support by the
Israel Science Foundation --- the Jack Adler Foundation for Space Research,
grant 63/01-1. A.V.F. is grateful for the financial support of NSF grants
AST 94-17213 and AST 99-87438, as well as of the Guggenheim 
Foundation.

\clearpage
\begin{deluxetable}{clccc}
\tabletypesize{\scriptsize}
\tablecaption{SN Spectral Database \label{database table}}
\tablewidth{17cm}
\tablehead{
\colhead{Type} & \colhead{SNe} & \colhead{Epochs} & \colhead{Redshift}&
\colhead{References\tablenotemark{a}} 
}
\startdata
Ia   & 1994D  & 22 & 0.0015 & 1,2 \\
     & 1987L  & 2  & 0.0074 & 1\\
     & 1995D  & 4  & 0.0066 & 2\\
     & 1999dk & 5  & 0.0150 & 2\\
     & 1999ee & 12 & 0.0114 & 3\\
Ib   & 1984L  & 12 & 0.0051& 1,2 \\
     & 1991ar & 1 & 0.0152 & 4 \\
     & 1998dt & 2 & 0.0150 & 4 \\
     & 1999di & 1 & 0.0164 & 4 \\
     & 1999dn & 3 & 0.0093 & 4 \\
Ic   & 1994I  & 14 & 0.0015 & 1,5 \\
     & 1990U  & 8 & 0.0079 & 4 \\
     & 1990B  & 4 & 0.0075 & 4 \\
II-P & 1999em & 27 & 0.0024 & 6\\
     & 1992H  & 13 & 0.0060 & 1,2\\
     & 2001X  & 12 & 0.0049 & 2,7\\
IIn  & 1998S  & 13 & 0.0030 & 8,9\\
     & 1994Y  &  1 & 0.0080 & 1 \\
     & 1994ak &  1 & 0.0085 & 1 \\
IIb  & 1993J  & 12 & 0 & 1 \\
     & 1996cb &  3 & 0.0024 & 10 \\
\tableline
Ic (98bw-like)& 1998bw & 25 & 0.0085 & 11\\
     & 2002ap & 19 & 0.0022 & 12,13\\
\tableline
Total & & 216 &&\\
\enddata
\tablecomments{Data from Poznanski et al. (2002) listed in the
upper part of the table. 
Spectra of SN 1998bw and SN 2002ap added in work study are listed
below. No extinction corrections were applied. 
SNe studied by Poznanski et al. (2002) were selected to avoid
highly extinguished events. Low extinction values are also
reported for SN 1998bw ($A_V\le0.2$; Patat et al. 2001) 
and for SN 2002ap ($A_V\le0.1$; Takada-Hidai, Aoki, \& Zhao, 2002).}
\tablenotetext{a}{
(1) Filippenko 1997; (2) Unpublished spectra by Filippenko and collaborators, 
obtained and reduced as those presented in (1), (4)--(6) and (8); 
(3) Hamuy et al. 2002; (4) Matheson et al. 2001; (5) Filippenko et al. 1995b;
(6) Leonard et al. 2002; (7) Gal-Yam \& Shemmer
2001; (8) Leonard et al. 2000; (9) Fassia et al. 2001; (10) Qiu et al. 1999;
(11) Patat et al. 2001; (12) Gal-Yam, Ofek, \& Shemmer 2002; (13)
Foley et al. 2003.\\}
\end{deluxetable}


\end{document}